# A monolithically integrated polarization entangled photon pair source on a silicon chip

Nobuyuki Matsuda[1,3], Hanna Le Jeannic[1,¶], Hiroshi Fukuda[2,3], Tai Tsuchizawa[2,3], William John Munro[1], Kaoru Shimizu[1], Koji Yamada[2,3], Yasuhiro Tokura[1,†], Hiroki Takesue[1]

[1]NTT Basic Research Laboratories, NTT Corporation, Atsugi, Kanagawa 243-0198, Japan

[2]NTT Microsystem Integration Laboratories, NTT Corporation, Atsugi, Kanagawa 243-0198, Japan

[3]Nanophotonics Center, NTT Corporation, Atsugi, Kanagawa 243-0198, Japan

[¶]Present address: ESPCI ParisTech, 10 rue Vauquelin, 75005 Paris, France

[†]Present address: Graduate School of Pure and Applied Science, University of Tsukuba, Tsukuba, Ibaraki 305-8571, Japan

**Integrated photonic circuits are one of the most promising platforms for large-scale photonic quantum information systems due to their small physical size and stable interferometers with near-perfect lateral-mode overlaps. Since many quantum information protocols are based on qubits defined by the polarization of photons, we must develop integrated building blocks to generate, manipulate, and measure the polarization-encoded quantum state on a chip. The generation unit is particularly important. Here we show the first integrated polarization-entangled photon pair source on a chip. We have implemented the source as a simple and stable silicon-on-insulator photonic circuit that generates an entangled state with 91 ± 2% fidelity. The source is equipped with versatile interfaces for silica-on-silicon or other types of waveguide platforms that accommodate the polarization manipulation and projection devices as well as pump light sources. Therefore, we are ready for the full-scale implementation of photonic quantum information systems on a chip.**

Photonic quantum states encoded in polarizations, paths, and times of arrival are excellent qubit candidates. Qubits are the quintessential resource for photonic quantum information processing (QIP), which provides a fundamentally new approach to communication [1], metrology [2], simulation [3-5], and computation [6-8]. Of these quantum states the polarization-encoded state is a true two-level photonic system, which is easy to manipulate with bulk optics such as waveplates. It is at the heart of many photonic QIP protocols [1, 4-9].

Integrated quantum photonics, which exploits miniature physical size and stable interferometers with near-perfect lateral-mode overlaps of integrated lightwave circuits, constitutes the future for realizing a scalable photonic QIP system on a chip [10, 11]. Recent experiments on integrated quantum photonics have utilized path-encoded quantum states of photons [3, 10-13]. However, polarization encoding allows us to implement the systems without the need for path duplication, and thus provides the simplest and most compact circuitry. It will also allow us to implement a wealth of the QIP protocols. To accomplish such polarization-encoded QIP systems, it is essential to develop integrated subsystems to generate, manipulate, and measure polarization-encoded quantum states on a single chip. Although the integration of optical circuits makes the handling of polarization states slightly more challenging, Bonneau *et al.* [14] and Sansoni *et al.* [15,16] recently demonstrated the manipulation and projection of quantum states on a chip by using a lithium niobate (LN) waveguide modulator and low-birefringent silica waveguides, respectively. An integrated single-photon detector on a silicon chip has already been demonstrated [17]. Therefore the only remaining task is to implement a polarization-entangled photon pair source [18,19]. The source is an essential component of such polarization-encoded QIP systems as quantum gates [7], on-demand single photon sources [20], and scalable one-way quantum computation [8]. Furthermore, path-polarization hybrid integration is also useful as a simulation tool for physical quantum systems [4].

Many polarizatio-entanglement sources have already been realized by means of integrated nonlinear waveguides, from which we can obtain quantum-correlated photon pairs via $\chi^{(2)}$ or $\chi^{(3)}$ spontaneous parametric processes [19, 21-25]. Such waveguides generally exhibit a polarization-mode walk-off, which causes the generated polarization entanglement to degrade. To compensate for the longitudinal walk-off, some of the sources required off-chip quantum erasers, namely an additional birefringence crystal, a polarization-dependent delay line, or a polarization-maintained fibre [21-23]. The others utilized one polarization mode of the nonlinear waveguides to avoid the walk-off, by using additional fibre-optic circuits for polarization steering [24-26]. Therefore, there have been no fully integrated polarization entanglement sources that generate photon pairs entangled inside the chip; nevertheless the feature is necessary for a practical integrated QIP system [27]. On the other hand, current optical communication technologies require sophisticated polarization diversity technologies [28, 29], which potentially solve the problem of integrating the polarization entanglement source.

In this work, we experimentally demonstrate the first polarization entanglement source that is fully integrated as a silicon photonic circuit. We compensate for the polarization-dependent walk-off simply by designing the device to be symmetric as regards the polarization degree with respect to the device midpoint, with the help of a polarization manipulation technology for telecommunication devices. The symmetric structure is also useful for the stable generation of maximally entangled states, in the presence of waveguide loss. Our source is not based on a post-selection scheme [22], so creates no spurious photons that limit the application of the source [21]. Furthermore, the device is equipped with a spot-size converter (SSC), which is an interface with silica or other types of quantum waveguide circuits for a fully integrated QIP system on a chip.

## Results

**Device design and building blocks.** Our source (Fig. 1), which is based on a simple and single-path configuration, consists of two silicon-wire-waveguide photon pair sources [30, 31] connected by an ultra-small silicon-wire polarization rotator [32]. Both silicon wire waveguides are 400 nm wide, 200 nm high and 1.5 mm long. The silicon polarization rotator has an off-axis double core structure. The inner core is a silicon photonic wire with a 200-nm-square cross section, which is embedded in a $SiO_xN_y$ outer core with an 840-nm-square cross section. The silicon wire waveguides and the silicon polarization rotator are connected by 10-µm-long tapered silicon wires. The over- and under-cladding of the silicon wire waveguides and the silicon polarization rotator are made of $SiO_2$ (not shown). The device is equipped with SSCs with tapered silicon cores at their ends [33]. In this work, the SSCs provide efficient coupling between the silicon wire waveguides and external optics. As we discuss later in detail, the SSC can be an interface between our source and silica-on-silicon or other types of waveguide platforms [34] for the integrated polarization-encoded QIP system.

In a silicon wire waveguide, a correlated pair of signal and idler photons are created via $\chi^{(3)}$ spontaneous four-wave mixing (FWM) following the annihilation of two pump photons [30, 31]. The spontaneous FWM in the silicon wire waveguide is useful for the efficient creation of photon pairs, thanks to the high $\chi^{(3)}$ nonlinearity of silicon in the telecom band (more than 100 times higher than that of silica) and the lateral confinement of light in a region less than a wavelength in size. The Raman photons created in the single-crystalline silicon core can be spectrally separated from the signal and the idler photons to reduce potential noise [30, 31]. Moreover, the pump and the photon pair have similar frequencies in the spontaneous FWM process, so the longitudinal walk off between involved photons becomes much smaller than those in the $\chi^{(2)}$ process [19]. Thanks to these features we can obtain highly pure photon pairs from a silicon wire waveguide with a length of few millimetres with a high efficiency.

In the silicon wire waveguide, TE-polarized photon pairs are efficiently generated by a TE-polarized pump field. Figure 2a shows spectra of the idler-to-signal conversion efficiency (extracted from the data shown in Supplementary Fig. S1 online) via the stimulated FWM for cases where the pump, signal, and idler fields are all TE-like (*i.e.*, horizontally polarized) and TM-like (*i.e.*, vertically polarized) (see Methods for details). The efficiency is proportional to the photon-pair production rate in spontaneous FWM [31]. The notably high FWM efficiency and its large bandwidth under the all-TE condition are originating from strong field confinement in the core (Supplementary Fig. S2 online) and the near-zero-dispersion property of the TE mode in the telecom band (Supplementary Fig. S3 online). Thus, we can regard our silicon wire waveguide as a photon pair source that operates only in the TE mode. In the present work, we utilize this polarization-dependent feature to build a polarization entanglement source.

The silicon polarization rotator was originally developed as a polarization-diversity integrated optical component for photonic networks [29]. The off-axis double core structure of the silicon polarization rotator exhibits two orthogonal eigenmodes, which have different effective refractive indices and eigen-axes tilted at 45° to the normal with respect to the silicon substrate (Fig. 2b). The birefringence in the eigenmodes provides an integrated waveplate, causing the polarization plane to rotate by an amount that depends on the length of the rotator. The insertion loss of the silicon polarization rotator was estimated to be approximately 1 dB [32]. We investigated the polarization rotation angle at the rotator $\theta_{SPR}$ (see Methods). Figure 2c shows the light transmission properties of the reference 3-mm silicon wire waveguide and the polarization entanglement source that includes a 30-µm-long silicon polarization rotator as a function of an additional output polarizer angle. A $\theta_{SPR}$ value of 86.7 ± 0.1° was obtained (from a fitting) with a fringe visibility $V$ of 0.99 ± 0.01, while the reference silicon wire waveguide exhibited a small initial offset of − 0.4 ± 0.3° with $V$ = 0.98 ± 0.01. Thus, a polarization rotation angle of almost 90° was obtained with the high polarization extinction ratio.

**Operating principle of the device for the stable single-chip generation of polarization-entangled photons.** We use a pump beam with + 45° (D) linear polarization, which is a 1:1 combination of the TE and TM modes in the silicon wire waveguides (Fig. 1). In the first silicon wire waveguide (SWW1), the TE component of the pump creates a photon pair in the $|TE,TE>_{s,i}$ state, which the silicon polarization rotator rotates to the $|TM,TM>_{s,i}$ state as a result of a 90° polarization rotation. Here the subscripts denote the frequency modes of the signal and idler photons. At the same time, the silicon polarization rotator rotates the TM component of the pump field to provide it with TE polarization, and the second silicon wire waveguide (SWW2) creates other |TE,TE⟩$_{s,i}$ photons. Since we cannot distinguish whether the pair was generated in SWW1 or SWW2, we obtain the maximally polarization-entangled state:

$$|\psi\rangle = (|TE,TE\rangle_{s,i} + e^{-i\phi}|TM,TM\rangle_{s,i})/\sqrt{2}, \qquad (1)$$

at the output of the polarization-entanglement source. It should be noted that our scheme is post-selection-free and thereby creates no spurious photons [22]. Here the relative phase difference $\phi$ depends on the length difference $\Delta L$ between SWW1 and SWW2. Strong birefringence in a silicon wire waveguide causes significant polarization-mode dispersion of photonic wave packets. In our silicon wire waveguide the estimated polarization-mode dispersion is 5 ps/mm, which is comparable to the typical width of a biphoton wave packet obtained from a silicon wire waveguide [30, 31]. However, since the polarization-entanglement source is designed to be symmetric as regards the polarization states with respect to the midpoint of the device, the polarization-mode dispersion of the pump pulses and photon pairs are cancelled out if $\Delta L = 0$. The $\Delta L$ fabrication error is less than 1 μm, which corresponds to a negligible walk off of less than 10 fs. Therefore, our polarization-entanglement source requires no additional component outside the chip to compensate for the temporal distinguishability of the biphoton wave packets caused by polarization-mode dispersion as in the earlier polarization-entanglement sources [21-23].

Integrated waveguides generally exhibit linear propagation loss, which reduces number of pairs created in SWW1, can cause the imbalance between the two terms in the entangled state. However, our configuration automatically equalizes the two amplitudes and thus provides a maximally entangled state. This is because the reduction of the pairs created in SWW1 resulting from the waveguide loss in SWW2 is equivalent to the reduction of the pairs to be created in SWW2 owing to the loss of the TM component of the input pump pulses. We denote the light transmittance of each silicon wire waveguide for the TM mode and the transmittance of the silicon polarization rotator as $\mu_{TM}$ and $\mu_{SPR}$, respectively. The photon pairs created in SWW1 by TE-polarized pump pulses suffer from extra loss in the silicon polarization rotator and SWW2 compared with the pairs created in SWW2. Hence, the output photon pair rate decreases by $(\mu_{TM}\mu_{SPR})^2$. On the other hand, the intensity of the TM-polarized component of the pump pulses decreases by $\mu_{TM}\mu_{SPR}$ in the SWW1 and the silicon polarization rotator. Thus, the pair production rate decreases by $(\mu_{TM}\mu_{SPR})^2$, which corresponds to the total transmittance of the other pair. Therefore, the two terms in the $|\psi\rangle$ state are automatically balanced. Note that this effect is not available with the $\chi^{(2)}$-based parametric process where the pair production rate is proportional to the pump power [18, 19].

**Observation of polarization-entangled photon pairs.** We have created polarization-entangled photon pairs using optical pulses with a temporal width of 80 ps and a repetition rate of 100 MHz in the experimental set-up shown in Fig. 3 (see Methods for details). The detected coincidence rate $N_{cc}$ and the coincidence over accidental ratio (CAR) of the photon pairs generated with the set-up were 103.9 Hz and 54.7, respectively, for the reference silicon wire waveguide with a projection basis of $|TE,TE\rangle_{s,i}$, and a pump peak power $P_{peak}$ of 69 mW. Given the device length of 3 mm and the overall photon detection efficiency of − 15 dB (per channel), the pair creation efficiency at the chip end was estimated to be 0.14 pairs/pulse/GHz/W$^2$/cm$^2$, which is comparable to the efficiency of 0.2 pairs/pulse/GHz/W$^2$/cm$^2$ in conventional silicon wire waveguides fabricated without a SiO$_x$N$_y$ deposition process [31]. At the same time, our polarization-entanglement source emitted photon pairs with $N_{cc}$ = 39.0 Hz and CAR = 57.9 with a $|TE,TE\rangle_{s,i}$ basis and $N_{cc}$ = 40.2 Hz and CAR = 42.2 at $|TM,TM\rangle_{s,i}$, at $P_{peak}$ of 128 mW. The slightly degraded CAR of the TM pairs compared with that of TE pairs is because the TM pairs (generated in SWW1) suffered additional loss in the silicon polarization rotator and SWW2.

We performed quantum state tomography on the generated photon pairs by carrying out a correlation measurement of 16 polarization combinations that were selected based on the angles of the four wave plates in the polarization projection units [35]. The reconstructed density matrix with the maximum-likelihood estimation for the reference silicon wire waveguide $\rho_{ref}$ and for the polarization-entanglement source $\rho_{ent}$ (with measurement times of 60 and 120 s) are shown in Fig. 4a and b, respectively, as their real and imaginary parts, and absolute values of each element. Note that these results included statistical accidental coincidence counts. $\rho_{ref}$ shows that the reference waveguide generated $|TE,TE\rangle_{s,i}$ pairs as expected. For the state $|\psi\rangle$ we obtain a density matrix written as

$$\rho = \sum |\psi\rangle\langle\psi| = \frac{1}{2}(|\text{TE, TE}\rangle\langle\text{TE, TE}| + e^{i\phi}|\text{TE, TE}\rangle\langle\text{TM, TM}| \\ + e^{-i\phi}|\text{TM, TM}\rangle\langle\text{TE, TE}| + |\text{TM, TM}\rangle\langle\text{TM, TM}|)$$ , (2)

if the generated state is purely $|\psi\rangle$. The off-diagonal terms, which represent the state purity, vanish if the biphoton is in a mixed state. Therefore, the state amplitudes (absolute values) close to 0.5 at the off-diagonal corners of the obtained $\rho_{ent}$ represent the high state purity. We obtained a high purity value of $0.93 \pm 0.02$.

As regards the degree of entanglement, we estimate the fully entangled fraction $F(\rho) = \max_{\Psi}\langle\Psi|\rho|\Psi\rangle$, where the maximum is taken over all maximally entangled states $|\Psi\rangle$, *i.e.*, over $|\Psi\rangle = U_s \otimes U_i \left[\frac{1}{\sqrt{2}}\left(|\text{TE, TE}\rangle_{s,i} + |\text{TM, TM}\rangle_{s,i}\right)\right]$, where $U_s$ and $U_i$ are unitary transformations on the signal and idler modes [36, 37]. Therefore, we can create any maximally entangled state (including Bell states) from a state $\rho$ with $F(\rho) = 1$ by employing linear optics such as wave plates. In accordance with the procedure described in ref [37], we obtained $F(\rho_{ref})$ as $0.51 \pm 0.02$, which is on the bound of the classical states of $F(\rho) = 0.5$. Thus, the single straight silicon wire waveguide created no entanglement. At the same time, for the polarization-entanglement source we obtained $F(\rho_{ent}) = 0.91 \pm 0.02$. The $F(\rho_{ent})$ value is much greater than $1/\text{Sqrt}(2) \sim 0.71$, implying that the generated state can violate the Clauser-Horne-Shimony-Holt inequality [38]. In addition, the concurrence (an alternate measure of entanglement) was obtained as $0.88 \pm 0.02$. Hence, we successfully generated photon pairs with a high degree of polarization entanglement from the chip.

## Discussion

The imperfection of $F(\rho_{ent})$ value is considered to be mainly due to the wavelength-dependent polarization rotation at the SSCs and at the silicon polarization rotator (see Supplementary for details). The unexpected polarization rotation at the SSCs might be due to a fabrication error (*e.g.*, a slight geometric horizontal offset between the axes of the inverse-taper silicon wire and the second core at the SSCs). Regarding the silicon polarization rotator, in principle, the retardation dispersion is $\Delta\theta_{SPR}/\Delta\lambda = 0.2°/\text{nm}$ [32], which is as small as usual bulk zero-order half wave plates [39]. This small dispersion is because the silicon polarization rotator is also a zero-order rotator. Therefore, by solving practical fabrication issues (such as the semi-circular shape of the $SiO_xN_y$ second core [32]) we could provide a better silicon polarization rotator with almost perfect polarization rotation.

To summarize, we have demonstrated a monolithically integrated polarization entangled photon pair source. The device requires no additional off-chip components and thus is the first complete subsystem of an integrated polarization entanglement source. To achieve a full-scale QIP syste, a polarization entanglement source must be capable of integration with quantum circuits consisting of manipulation and projection devices for polarization-encoded states on the same chip. Furthermore, it is argued that the polarization entanglement source should also be equipped with an electrically driven pump laser integrated on the same chip for further system stability [40]. These criteria seem to be formidable at first glance; however, our source can be integrated with any of them. The silica SSC interface allows hybrid integration with silica-on-silicon waveguides [34], which already accommodates a polarization controller [41], a polarization beam splitter [16, 42], and even a pump laser diode [43]. Although a rectangular-core silica waveguide generally exhibits a strain that makes the waveguide birefringent [16], such birefringence can be eliminated [42] or tuned with voltage-controlled heaters [44]; the birefringence compensation technology is useful for, *e.g.*, selecting output entangled states on a chip by tuning the phase difference $\phi$ in the state $|\psi>$. Furthermore, the silica waveguides can be integrated with high-speed LN-based modulators [14] by means of silica-LN hybrid integration technology [45]. Of course our silicon-wire-based device can be directly connected with silicon-wire-based quantum circuits [13]. Integrated single-photon detectors on a silicon chip are also ready with a markedly high quantum efficiency [17]. Therefore, our monolithic polarization-entangled photon pair source helps pave the way to the full-scale implementation of photonic QIP system using the polarization degree of freedom.

## METHODS

### Device fabrication

Our device was fabricated on a silicon-on-insulator wafer with a 200-nm-thick silicon layer and a 3-µm-thick buried $SiO_2$ layer. The silicon wire waveguides were 400 nm wide. The 30-µm-long and 200-nm-wide silicon wire for the rotator was adiabatically connected by 10-µm-long silicon tapers. For efficient coupling between the device and the external fibres, we equipped both ends of the device with SSCs. The silicon wire waveguides and SSCs were fabricated by electron beam lithography and electron cyclotron resonance plasma etching. An 840-nm-thick silicon oxynitride film with a refractive index of 1.60 was deposited by plasma-enhanced chemical vapour deposition and the second core of the silicon polarization rotator was formed by reactive ion etching with fluoride gas. The measured propagation losses of our device were 2.2 and 1.7 dB/cm for the TE and TM modes, respectively.

### Details of stimulated FWM experiment

In the stimulated FWM experiment designed to show the discrepancy between the FWM efficiencies of the TE and TM polarizations, we used two independent wavelength-tuneable cw lasers as a pump and a signal source. The pump beam was modulated into a pulse train (500-ps full-width half maximum, 40-MHz repetition rate) with an LN intensity modulator. The pump and signal beams were combined by a 50/50 directional coupler, and then co-polarized in either the TE or TM mode by the half-wave plate in the free-space focusing module, which coupled the light into a straight silicon wire waveguide with a length $L$ of 3 mm that was used as a reference. They were then corrected from the waveguide output with a lensed fibre, and the overall output spectrum was measured directly with an optical spectrum analyzer (OSA) with a spectral resolution of 0.05 nm. The observed FWM spectrum is shown as a density plot in supplementary Fig. S1. Here the signal wavelength was scanned across the telecom band to change the pump-signal detuning, while the pump wavelength was fixed at 1551.1 nm. The pump and the signal powers coupled to the waveguide were 90 mW (peak) and 3.0 mW, respectively. The data in Fig. 2a were extracted from the data in Fig. S1. Here the stimulated FWM efficiency, which is the ratio of the output idler power to the output signal power, is approximated as $\eta = (\gamma P_{peak} L)^2$ in the vicinity of zero pump-signal detuning [31]; the nonlinear constant $\gamma = (2\pi n_2)/(\lambda_p A_{eff})$, where $n_2$ is the nonlinear refractive index of the material, $\lambda_p$ is the pump wavelength. $A_{eff}$ is the polarization-dependent effective mode area, which mainly determines the discrepancy between the $\eta$ values in the TE and TM modes. The calculated $A_{eff}$ ratio between the TE and TM modes is 0.2, corresponding to an $\eta$ ratio of $0.2^2$ (− 14 dB), which approximately explains the experimental value of − 17 dB.

**Evaluating the performance of the silicon polarization rotator**

In Fig. 3, IM consists of an electro-absorption modulator and an LN intensity modulator, which modulated a continuous beam from the light source ($\lambda_p$ = 1551.1 nm) into a train of pump pulses. The pulses were amplified by an erbium-doped fibre amplifier, filtered to eliminate the amplified spontaneous emission noise, and then launched into the polarization-entanglement source. The input polarization state was set by using a free-space focusing module that contained a collimation lens followed by a polarizer, a half-wave plate, and a high-NA focusing lens. Here the coupling efficiency with the chip was − 2 dB.

To evaluate the polarization rotation angle of the silicon polarization rotator, we set the input half-wave plate angle at 0° so that the input state was in the TE mode. We also used a similar free-space focusing module but with an embedded polarizer to collect and analyse the SOP of the output field. Figure 2b is the measured transmittance as a function of the output polarizer angle.

**Experimental details of the polarization entanglement generation**

For the polarization-entanglement generation, we set the half-wave plate in the input focusing module at + 22.5° so that the pump pulses were D polarized. Then the output field from the device including entangled photons was collected by a lensed fibre with an outcoupling efficiency of − 2 dB. Then, the light was introduced into the WDM filter, which suppressed the residual pump field (with an isolation of more than 130 dB) and separated the signal and idler photons into different fibre channels. Each output port has a centre wavelength of 1546.4 nm (s) and 1556.0 nm (i) with a channel bandwidth of 0.14 nm (18 GHz). Then, the photons passed through fibre birefringence compensators (a half wave plate sandwiched by two quarter wave plates) and subsequently polarization analysers, each of which consisted of a half and a quarter wave plate, and a polarizer. Finally, the photons were received by InGaAs single photon detectors (id210, id Quantique) that operated at a gate frequency of 100 MHz synchronous with the pump's repetition rate. The quantum efficiency, gate width, the dark count rate, and the dead time of the detectors were 20 %, 1 ns, $10^{-5}$ /gate, and 5.0 μs, respectively. The overall photon detection efficiency was − 15 dB per channel, which consisted of the collection efficiency of the created photons (− 8 dB) and the quantum efficiency of each detector (− 7 dB). The detection signals from the two detectors were input into the time interval analyser (TIA) for coincidence measurements.

Each wave plate of the birefringence compensators was set so that the polarization basis at the device output corresponded to the basis of the polarization analyser. To accomplish this, we used an erbium-doped fibre amplifier emitting an amplified-spontaneous-emission source in the TE mode at the input. Then, we set the angles of the wave plates in the birefringence compensators so that the minimum and maximum transmittance through the polarization analysers were obtained with appropriate basis sets.

The errors in the $F(\rho)$ values obtained with quantum state tomography were estimated using the angular setting uncertainties of the wave plates in the polarization projection units and the Poissonian noise statistics of the obtained coincidence counts. We increased the $P_{peak}$ value and the measurement time for the $\rho_{ent}$ measurement to obtain good statistics. The obtained total coincidence counts (the sum of the values on the diagonal bases) were $6.6 \times 10^3$ and $1.0 \times 10^4$ for the reference silicon wire waveguide and the entanglement source, respectively. The corresponding pair production rate was estimated to be $1.1 \times 10^{-3}$ and $0.8 \times 10^{-3}$ per pulse.

**Acknowledgements**

We are grateful to T. Watanabe, K. Azuma, M. Oguma, H. Takahashi, J. L. O'Brien, and S. Itabashi for fruitful discussions. This work was supported in part by a Grant-in-Aid for Scientific Research (No. 22360034) from the Japan Society for the Promotion of Science.

**Contributions**

NM and HL performed the experiments and data analysis. NM developed the device concept. HF, TT and KY designed and fabricated the sample. YT and HT led the project. All authors discussed the results and contributed to the preparation of the manuscript.

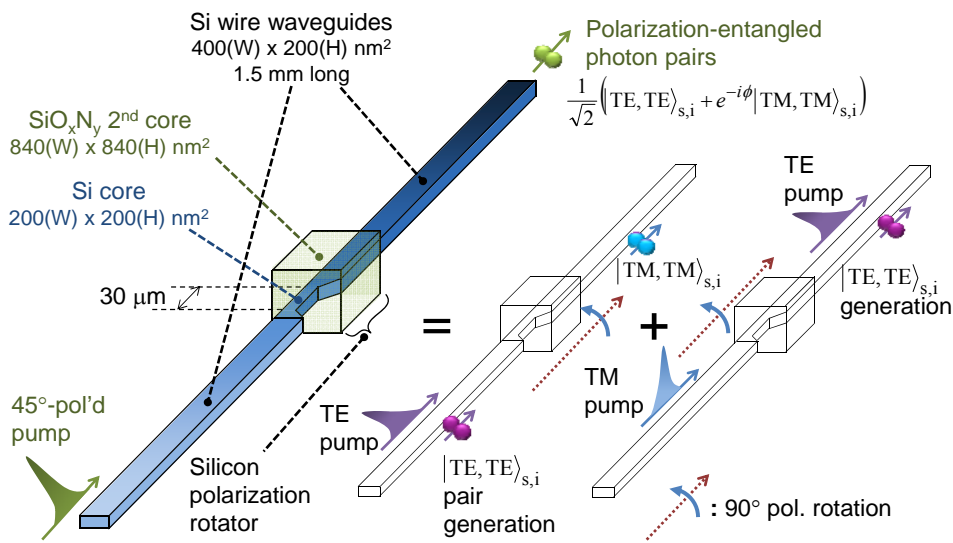

**Figure 1** | **A monolithically integrated polarization-entanglement source.** The source, fabricated on a silicon-on-insulator substrate, consists of a silicon-wire-based 90° polarization rotator sandwiched by two nonlinear silicon wire waveguides. The device generates the polarization entanglement as a superposition of the two events shown on the right hand side. The figure is not to scale for clarity.

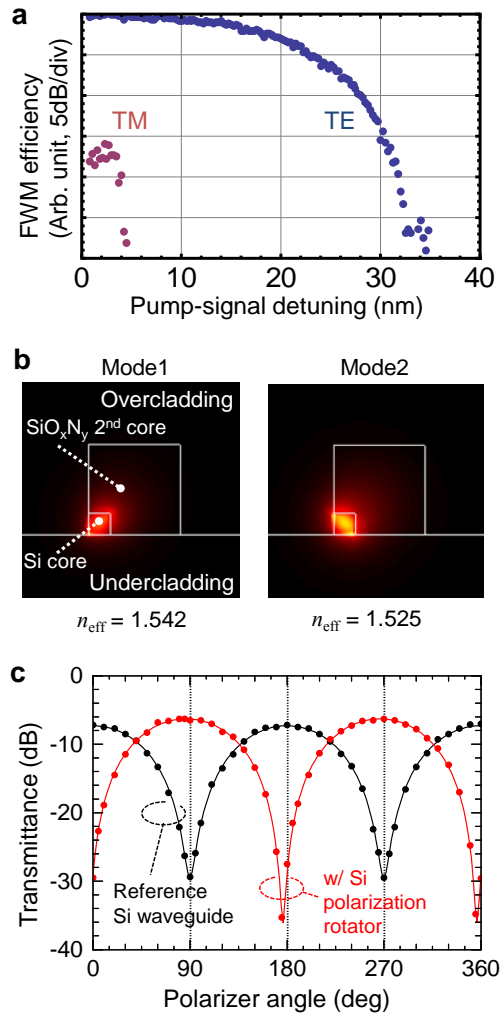

**Figure 2 | Property of each building block of the polarization-entanglement source. a.** Polarization dependence of the FWM efficiency in a silicon wire waveguide investigated with a stimulated FWM experiment. The idler-to-signal conversion efficiency is plotted a function of pump-to-signal detuning. The values are normalized with respect to the TE-FWM efficiency at zero detuning. **b.** The two eigenmodes in the silicon polarization rotator (simulated with a mode solver) with effective refractive indices $n_{\text{eff}}$. **c.** Measured transmittance of the silicon polarization rotator (red) and the reference silicon wire waveguide (black) with TE-polarized incident light as a function of output polarizer angle. The solid curves represent sinusoidal fittings.

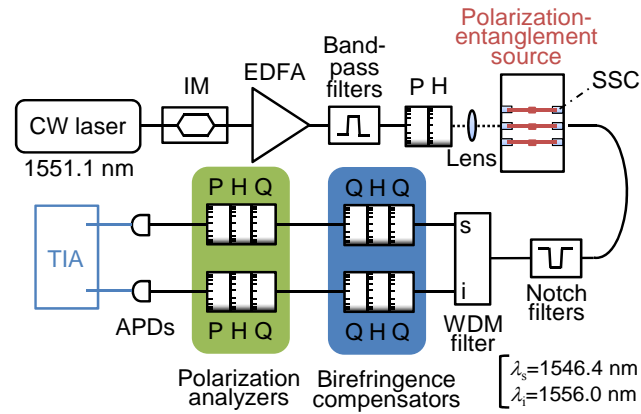

**Figure 3 | Experimental set-up for measuring the polarization entanglement.** IM, intensity modulator; EDFA, erbium-doped fibre amplifier; P, polarizer; H, half-wave plate; Q, quarter-wave plate; SSC, spot-size converter; APD, InGaAs avalanche photodiode; TIA, time-interval analyser. Dashed line shows free-space optical path. Blue lines show electrical connection. Note that all the wave plates are free space bulk optics embedded in collimation benches.

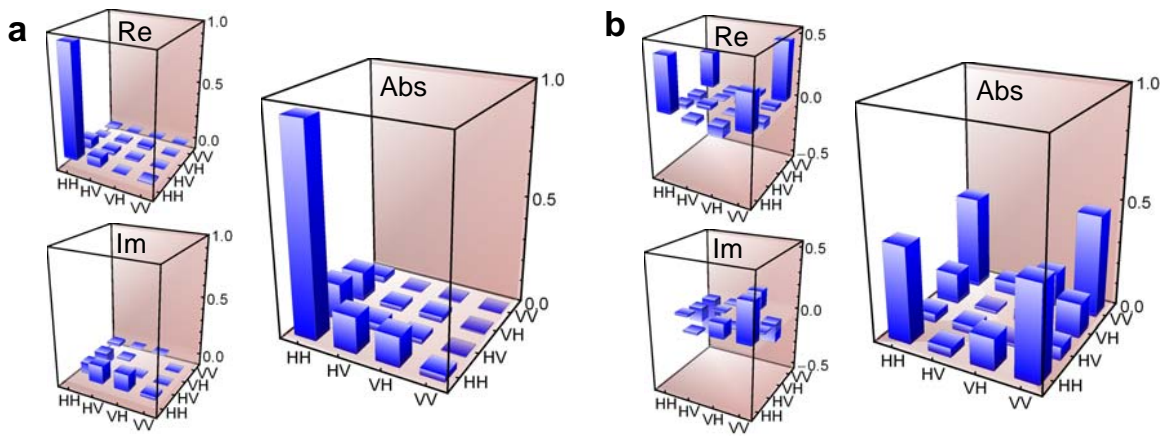

**Figure 4 | The real and the imaginary parts of the reconstructed density matrices for the two-photon polarization states generated from (a) the reference silicon wire waveguide and (b) the polarization entanglement source.** The absolute values of the elements are also displayed to show the amplitudes on off-diagonal elements. H and V represent the TE and TM bases, respectively.

## Supplementary information

### - Polarization dependence of FWM efficiency in a silicon-wire waveguide

In Fig. S1, we see an almost straight line consisting of created idler peaks, which appear almost symmetric with the signal line with respect to the pump wavelength. This is explained by the energy conservation of the FWM. The clear idler peak was observed only for the TE-polarized FWM, due to a strong field confinement in the core (Fig. S2) and the near-zero-dispersion property of the TE mode in the telecom band (Fig. S3). The data in Fig. 2a was extracted from the data in Fig. S1.

### - Imperfection in obtained fully entangled fraction $F(\rho_{ent})$

First, to verify the reproducibility of the experiment we measured $\rho_{ent}$ four times, and on each occasion we reset the input pump polarization manually using the input half-wave plate. The average value and the standard error were $F(\rho_{ent}) = 0.91 \pm 0.02$, which means that the angular settings of the wave plates were sufficiently reproducible. Second, the value of $F(\rho_{ent})$ without accidental coincidence counts was obtained as $0.92 \pm 0.02$, which indicates that statistical background coincidence events contribute little to the degraded fidelity.

The density matrix $\rho_{ref}$ in Fig. 4a exhibits $|TE,TM>_{s,i}$ and $|TM,TE>_{s,i}$ components, implying that the generated $|TE,TE>_{s,i}$ pairs in the reference silicon-wire waveguide suffered slight polarization rotation. We investigated this for the signal and idler wavelength modes using a classical amplified-spontaneous-emission source. The light source passed through the WDM filter before being coupled to the reference silicon-wire waveguide so that its spectrum was identical to that of the filtered photon pairs. Then, the polarization rotation property of the device was investigated (Fig. S4) in the same way as for the pump wavelength. As can be seen, there is a horizontal offset in the fringes for the signal and idler modes, even without the silicon polarization rotator. The unexpected polarization rotation might be due to a fabrication error, which is presumably a slight horizontal offset between the axis of the inverse-taper silicon wire and the second core at the SSCs. The amplified-spontaneous-emission source centred at the signal and the idler wavelengths had polarization rotations of $-11.0 \pm 0.2°$ and $-9.9 \pm 0.2°$, respectively. If these values are contributed by the SSCs at the input and the output equally, the created $|TE,TE>_{s,i}$ pairs in the reference SWW can be expected to suffer a polarization rotation of approximately − 6° at the output SSC. We plot the estimated density matrix for this case in Fig. S5, which well describes the feature of the experimentally obtained $\rho_{\text{ref}}$ (Fig. 4a). In addition we can also see the wavelength-dependent polarization rotation at the silicon polarization rotator.

These insufficient polarization rotations cause the propagating optical field to have a slightly rotated linearly polarized state from either the TE or TM mode that suffers from the strong birefringence of the following silicon-wire waveguides. As a result, the propagating field exhibits a rapid oscillation of the polarization state in the spectral region [32]. Such polarization oscillation is averaged out in the WDM filter window for the signal and idler modes, causing the degree of polarization to degrade. Therefore, we obtained the degraded fringe visibility seen in Fig. S4 and also an imperfect $F(\rho_{\text{ent}})$ value. We obtained a fringe visibility of 0.92. $F(\rho_{\text{ent}})$ could degrade to the same degree.

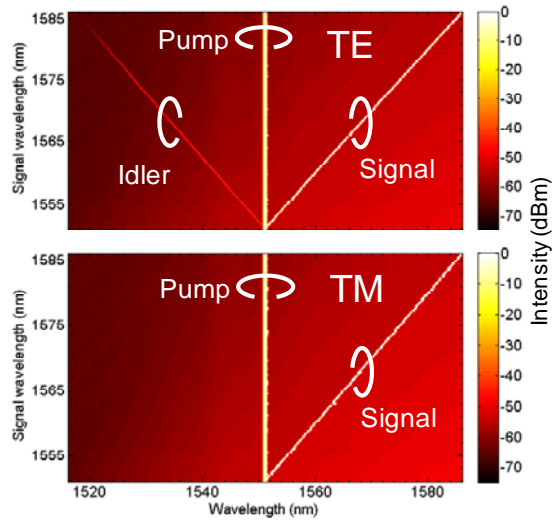

**Figure S1** Polarization dependence of the FWM efficiency in a silicon-wire waveguide investigated with a stimulated FWM experiment. Each density plot shows the observed FWM spectrum as a function of the signal wavelength (vertical axis).

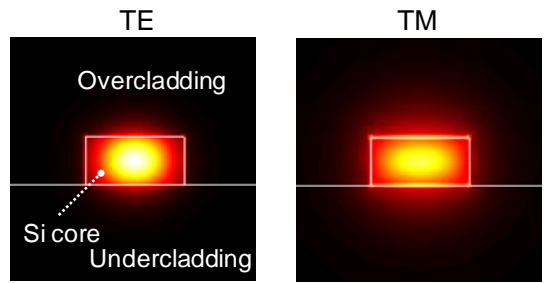

**Figure S2** Field distribution of the fundamental modes in the silicon-wire waveguide, calculated with a mode solver.

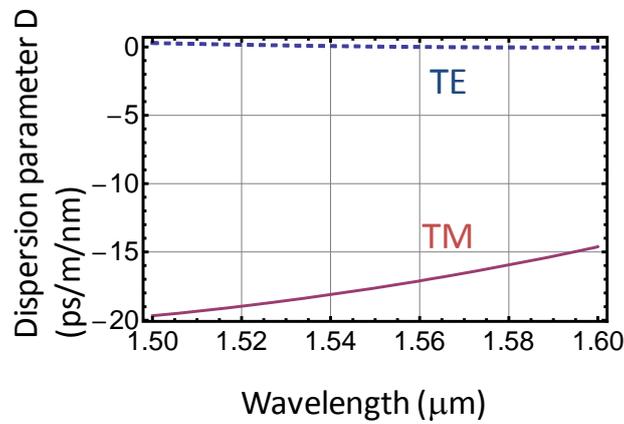

**Figure S3** Calculated polarization-dependent dispersion property in the silicon-wire waveguide.

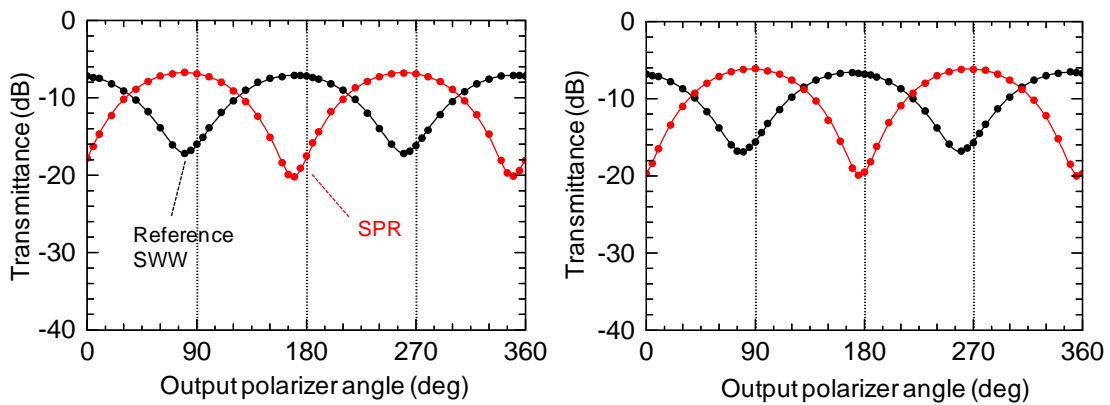

**Figure S4** Measured transmittance as a function of the output polarizer angle at the signal (left) and idler (right) wavelengths.

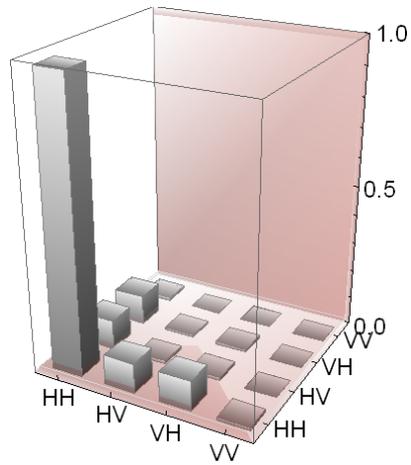

**Figure S5** Calculated absolute values of the elements of $\rho_{\text{ref}}$ in the presence of the finite polarization rotation at the output SSC. H and V represent the TE and TM modes, respectively.